# Estimating length of hospital stay, with regard to associated factors; A model to enhance healthcare service efficiency and reduce healthcare resource consumption

**Authors:**

1. **Seyed Nasser Moosavi (corresponding author)**

(Department of industrial engineering, Yazd university, Iran)

**Cell NO:** +989138216471

**Email:** seyednasserm6@gmail.com

2. **Ashkan Khalifeh**

(Department of Statistics, Yazd University, Iran)

**Cell NO:** +989134115217

**Email:** a.khalifeh@stu.yazd.ac.ir

3. **Ali Shojaee**

(Member of the board of directors of Iranian Health Insurance Organization (IHIO), Tehran, Iran – MD, MPH)

**Cell NO:** +989121307265

**Email:** shojaeeali1@yahoo.com

4. **Masoud Abessi**

(Professor of Information Systems, Department of industrial engineering, Yazd university)

**Cell NO:** +989132581706

**Email:** mabessi@gmail.com

**Abstract**

To assess the efficiency and the resource consumption level in healthcare scope, many economic and social factors have to be considered. An index which has recently been studied by the researchers, is "length of hospital stay" (LOS) defined as how long each patient is hospitalized. Detecting and controlling the factors affecting this index can help to reduce healthcare costs and also to improve healthcare service efficiency. The effects of three major factors, say, the season during which the patient is hospitalized, the patient's age and his/her gender on LOS pertaining to 82718 patients receiving healthcare service from the hospitals under contract to insurance organization in Tehran, have been analyzed, using unbalanced factorial design. The test results imply that the whole model is significant (P-value=0<0.01), the separate effects of all three factors on LOS are significant (P-value=0<0.01) and the only significant interaction at $\alpha=0.01$ is the one between gender and age group (P-value=0<0.01). moreover, no two age groups have the significant equal means and age groups 1-10 and 26-40 years possess the minimum and maximum means of LOS, respectively. LOS means in winter and autumn are equal, being the maximum. Due to the significance of these effects, allocating specified budget and resources with regard to these factors will help hospitals and healthcare centers enhance their efficiency and remain within their budget.

## 1. Introduction

A very important index used to determine healthcare service efficiency and healthcare resource consumption level is the patient's length of hospital stay or "length of stay" (LOS). This index is defined as length of the time elapsed for each patient since entering the hospital or healthcare center until discharge. Detecting and controlling the factors affecting LOS can help to reduce healthcare costs and also to improve healthcare service efficiency. These factors might be environmental, economic, technological, medical etc. So far, many researchers have studied different factors having association with LOS. Mitchell et al. [1] used a cohort study design to indicate that healthcare-associated urinary tract infection (HAUTI) is associated with extra length of hospital stay, a burden to the hospital system and required to be reduced by surveillance and interventions. Loudon et al. [2] proved that cardiovascular disease (CVD) factors including different indices have significant and varied impacts on LOS and mortality in patients with acute coronary syndromes. Gonzalez et al. [3] studied the impact of using the CLINITEK AUWi system on patients' LOS at a community teaching hospital. A negative association was found between them. Using regression models and matched subgroup analysis, Gross et al. [4] evaluated the contribution of practice variation (comorbidities, operative traits, and postoperative complications) to postoperative length of stay (pLOS) for children with perforated appendicitis. A significant variation in pLOS wasn't justified by the considered factors. Keswani et al. [5] studied the patients' LOS data as one of the primary outcomes of two treatment methods, endoscopic resection (ER) and surgical resection (SR) and offered to use these data when counseling patients about treatment options. Gärtner et al. [6] investigated whether Geriatric Nutritional Risk Index (GNRI) predicts hospital mortality, LOS and inflammatory markers. Two of these factors, say, LOS and inflammatory markers proved to correlate with GNRI. Studying hospitalized adult patients with diabetes along with community-acquired pneumonia (CAP), Bader et al. [7] indicated that delayed administration of congruous antibiotic therapy and moderate-to-severe pneumonia increase the risk of complication and LOS. Inneh [8] assessed the collective association of sociodemographic, preoperative comorbid and intraoperative factors with longer LOS after total knee arthroplasty. In a study for which the required data were compiled at a general hospital in Korea, Choi et al. [9] investigated how indoor daylight environments affected patients' overall length of stay (ALOS). In a retrospective cohort study, Wong et al. [10] indicated how LOS is affected by experiencing falls during inpatient stroke rehabilitation. The falls experienced contributed to a longer LOS. Carter et al. [11] demonstrated that psychiatric comorbidities have a significant and clinically important impact on LOS in heart failure patients in the UK. Gomes et al. [12] introduced risk of malnutrition as an independent predictor of mortality, LOS and hospitalization costs in stroke patients at six-months post stroke. Performing a multi-stage modeling for data relating to patients at a German university hospital, Arefian et al. [13] indicated that healthcare-associated infections result in extra length of stay and associated per diem cost. Garza-Ramos et al. [14] studied the patients over the age of 65 who underwent posterior surgery for cervical myelopathy, and indicated that hospital charge and mortality rate are profoundly higher for patients experiencing prolonged length of stay (PLOS) defined as stay beyond 6 days. Padegimas et al. [15] compared LOS after shoulder arthroplasty at an orthopedic specialty hospital (OSH) and a tertiary referral center (TRC). They showed that LOS at the OSH was significantly shorter than at the TRC. Performing a multivariate logistic regression analysis, Manoli et al. [16] also stated that among different operation methods for patients with proximal humerus fractures, total shoulder arthroplasty (TSA) leads to increased hospital costs despite a shorter LOS.

Despite the expansive research on LOS index and the factors associated with that, many other effective factors are yet to be well considered. In many hospitals and healthcare centers, although during some periods of the year, the need for extra facilities and staff to give an acceptable service is profoundly tangible, in other times, the hospital is overstaffed due to limited number of the patients. In the first case, the result is the patients' dissatisfaction and in the latter, the staff are overpaid. Finding a model that indicates the patients' LOS in different seasons can help to resolve the problem by allocating resources proportional to the hospitals' needs. The disease type is also another factor associated with LOS. Two factors, gender and age can affect the patient's disease type and consequently LOS. Regarding these points, we propose a statistical model that simultaneously considers the effects of three factors, "season" indicating when the patient is hospitalized, "gender" stating whether the patient is male or female and "age group" indicating how old the patient is, on the response variable LOS. The rest of the paper is organized as follows; section 2 explains the methods and materials used for this research. In section 3, we find the proper sample size resulting in an accurate model. The model's sufficiency is assessed in section 4. The results derived from the model are given in section 5, and finally section 6 concludes the paper.

## 2. Methods and materials

In this paper, we study the information relating to 82718 patients who have received healthcare service from the hospitals under contract to Salamat insurance organization, in Tehran, the capital of Iran. The factor "gender" states whether the patient receiving service is male or female. The factor "season" represents the season when each patient is hospitalized in the hospital or healthcare center to receive service and it has four levels. The factor "age group" states how old the patient is. We have categorized the patients into 5 major age groups. The first group consists of the patients aged 1 to 10, the second group consists of the patients 11 to 25 years old, the patients at the age of 26 to 40 form the third group, the fourth group is comprised of the patients 41 to 60 years old, and the patients aged over 60 are in the fifth group. We try to investigate the effects of three mentioned factors on variable LOS indicating how long (many days) each patient stays in the hospital. The unbalanced factorial design with three fixed-effect factors is applied to study the effects (because the number of replications in different cells is not the same and all levels of the factors are considered in the model). We have used IBM SPSS software (version 22) to process the data.

## 3. Choosing the sample size

In any design of experiment (DOE) problem, a crucial decision is how to choose the proper sample size (the number of replications in each experiment). The more we are concerned about small shifts in the values of LOS, the bigger the sample size must be. Since the length of time each patient stays in the hospital has a profound effect on healthcare costs and on a large scale, the nominal changes in the values of this variable profoundly affect the total cost, we are interested in detecting small shifts in LOS, caused by aforementioned factors. Hence, we increase the number of replications until the desired sensitivity is acquired. The operating characteristic (OC) curve is used to determine the proper number of replications. This curve plots the probability of type II error for different sample sizes against parameter Ø indicating how the null hypothesis doesn't hold. For each of the three factors introduced above, the null hypothesis states that the means of response variable at all levels of the factor are equal, and the alternative one states that at least two of the means aren't equal. Considering three current factors, we have seven formulas for $Ø^2$, three of which relate to the main effects and others are used for interactions. As an instance, the equation pertaining to the main effect of factor "season" is as follows.

$$Ø^2_{season} = \frac{ngaD^2}{2s\sigma^2} = 0.1328n \qquad (1)$$

where $D$ is a value that if the difference between two treatment means exceeds, the null hypothesis (the equality of treatment means) is rejected. $a, g$ and $s$ are the numbers of levels for factors "age group", "gender" and "season", respectively, and $n$ represents the required number of replications. We consider $D = 1$ day as a significant difference for each two means, resulting in rejection of the null hypothesis with probability at least 0.95 (the test power). To determine the variance value, first, we chose 100 replications and estimated $\sigma^2$ as $\widehat{\sigma^2} = s^2 = 9.41$. Now, choosing $\alpha = 0.01$ and with the aid of Fig. 1 indicating OC curve for numerator freedom degree (NFD) $\omega_1 = s - 1 = 4 - 1 = 3$ and denominator freedom degree (DFD) $\omega_2 = sga(n - 1) = 4 * 2 * 5 * (n - 1) = 40(n - 1)$ [17], we try to find the required sample size.

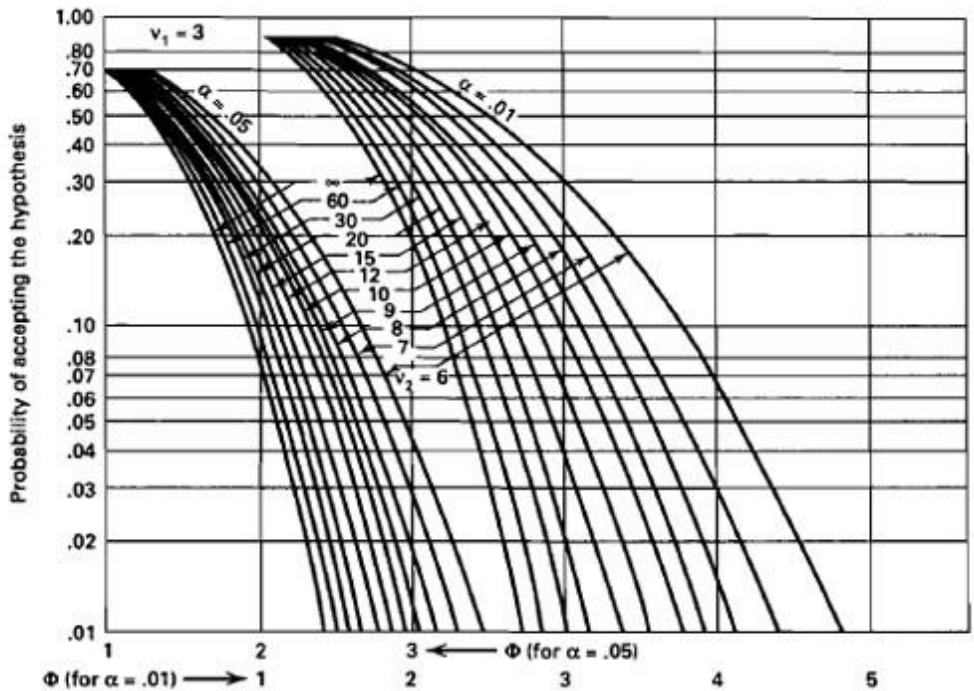

Figure 1. OC curve for the fixed-effects model analysis of variance ($\omega_1 = 3$)

As the first guess about the required sample size, we choose $n = 10$ replications, resulting in $\emptyset^2 = 1.328$ or $\emptyset = 1.1523$ and $DFD = 360$. Hence regarding Fig. 1, we have $\beta = 0.80$ and $\Pi = 1 - \beta = 0.20$ which is less than the desired value 0.95, implying that $n = 10$ replications isn't enough. Continuing this procedure, Table 1 is calculated.

Table 1. Calculating the sample size relating to the main effect of "season"

| n | Ø | numerator freedom degree(NFD) | denominator freedom degree(DFD) | β |
|---|---|---|---|---|
| 10 | 1.1523 | 3 | 40*(9)=360 | 0.80 |
| 20 | 1.6297 | 3 | 40*(19)=760 | 0.31 |
| 30 | 1.9959 | 3 | 40*(29)=1160 | 0.20 |
| 40 | 2.3047 | 3 | 40*(39)=1560 | 0.06 |
| 43 | 2.3896 | 3 | 40*(42)=1680 | 0.04 |

As shown in this table, with $n = 43$ replications, the probability of type II error is approximately 0.04 and the test power equals 0.96 which is even more than the desired value. It is interpreted as once the difference between two means relating to two levels of factor "season" is 1 day, the probability of rejecting the null hypothesis is 0.96, which concludes when the estimated value of variance is not seriously irrelevant, 43 replications is enough to get the required sensitivity and accuracy. Repeating this procedure for other main effects and interactions, we indicate that at least 480 replications is required to have an accurate and valid model. In this paper, an unbalanced three-factor design will be introduced in which the smallest number of replications pertains to cell (winter, female, age group 1) and equals 609, resulting in desired sensitivity. Table 2 indicates the number of patients in different levels of the factors.

Table 2. The frequency of the patients in different levels of the factors

| gender | | | age group | | | | | total |
|---|---|---|---|---|---|---|---|---|
| | | | 1 | 2 | 3 | 4 | 5 | |
| male | season | spring | 964 | 1120 | 1519 | 4073 | 4782 | 12458 |

|        |        |        | 1018 | 1369 | 1748  | 4083  | 3850  | 12068 |
|        |        | summer | 1018 | 1369 | 1748  | 4083  | 3850  | 12068 |
|        |        | autumn | 934  | 1235 | 1578  | 3710  | 3382  | 10839 |
|        |        | winter | 776  | 1233 | 1638  | 3856  | 3642  | 11145 |
|        | total  |        | 3692 | 4957 | 6483  | 15722 | 15656 | 46510 |
| female | season | spring | 735  | 644  | 993   | 3055  | 4078  | 9505  |
|        |        | summer | 718  | 868  | 1284  | 3215  | 3411  | 9496  |
|        |        | autumn | 679  | 673  | 1135  | 2941  | 3107  | 8535  |
|        |        | winter | 609  | 733  | 1169  | 2957  | 3204  | 8672  |
|        | total  |        | 2741 | 2918 | 4581  | 12168 | 13800 | 36208 |
| total  | season | spring | 1699 | 1764 | 2512  | 7128  | 8860  | 21963 |
|        |        | summer | 1736 | 2237 | 3032  | 7298  | 7261  | 21564 |
|        |        | autumn | 1613 | 1908 | 2713  | 6651  | 6489  | 19374 |
|        |        | winter | 1385 | 1966 | 2807  | 6813  | 6846  | 19817 |
|        | total  |        | 6433 | 7875 | 11064 | 27890 | 29456 | 82718 |

## 4. Testing and adjusting the model's sufficiency

To have a model that gives an accurate description of the data, the basic assumptions of factor analysis must hold, stating that the residuals are independent and normally distributed with mean zero and constant variance $\sigma^2$. Once these assumptions aren't proved in the model, the results derived from the model aren't valid. The model's validity is assessed by analyzing the residuals defined as $e_{ijkl} = y_{ijkl} - \hat{y}_{ijkl}$ ($i = 1, \ldots, g$, $j = 1,2, \ldots, s$, $k = 1,2, \ldots, a$).

### 4.1. Testing the normality assumption

The normality of residuals is tested by drawing the residuals' histogram. If the residuals come from a normal population, this chart must be bell-shaped. As Fig. 2 indicates, it is a little skewed to the right and the negative residuals aren't as large as we expect. Although a mediocre violation of normality is not worrisome, it might be possible to improve the model by adjusting the data.

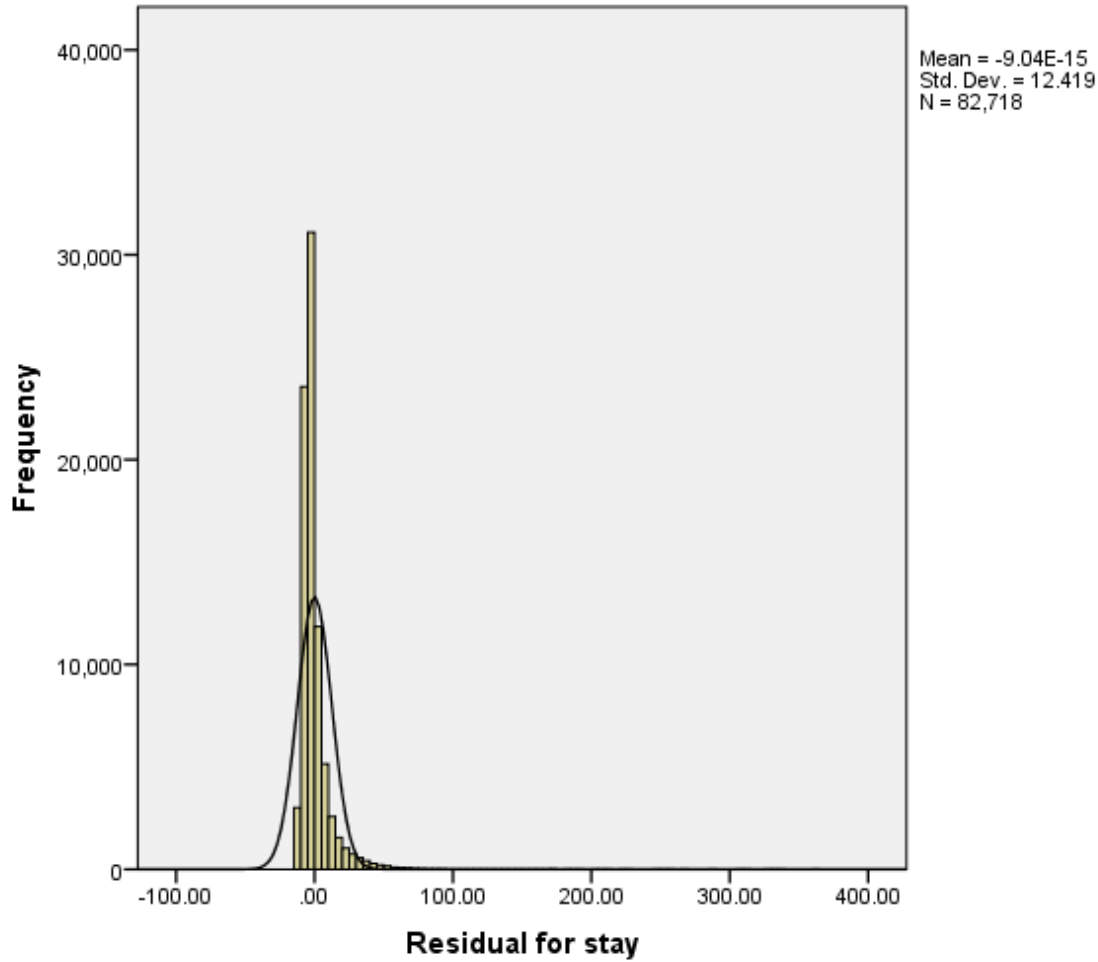

Figure 2. The residuals' histogram

## 4.2. Testing the variance constancy assumption

If the model is valid, the residuals lack any specific structure and do not tend to depend on other variables including the dependent (response) variable. Fig. 3 plots the residuals against the predicted values of the response variable. The residuals' variability has first increased, then decreased and again increased. The changing variance occurs in cases where the data follow a skewed normal distribution (as mentioned in section 4.1).

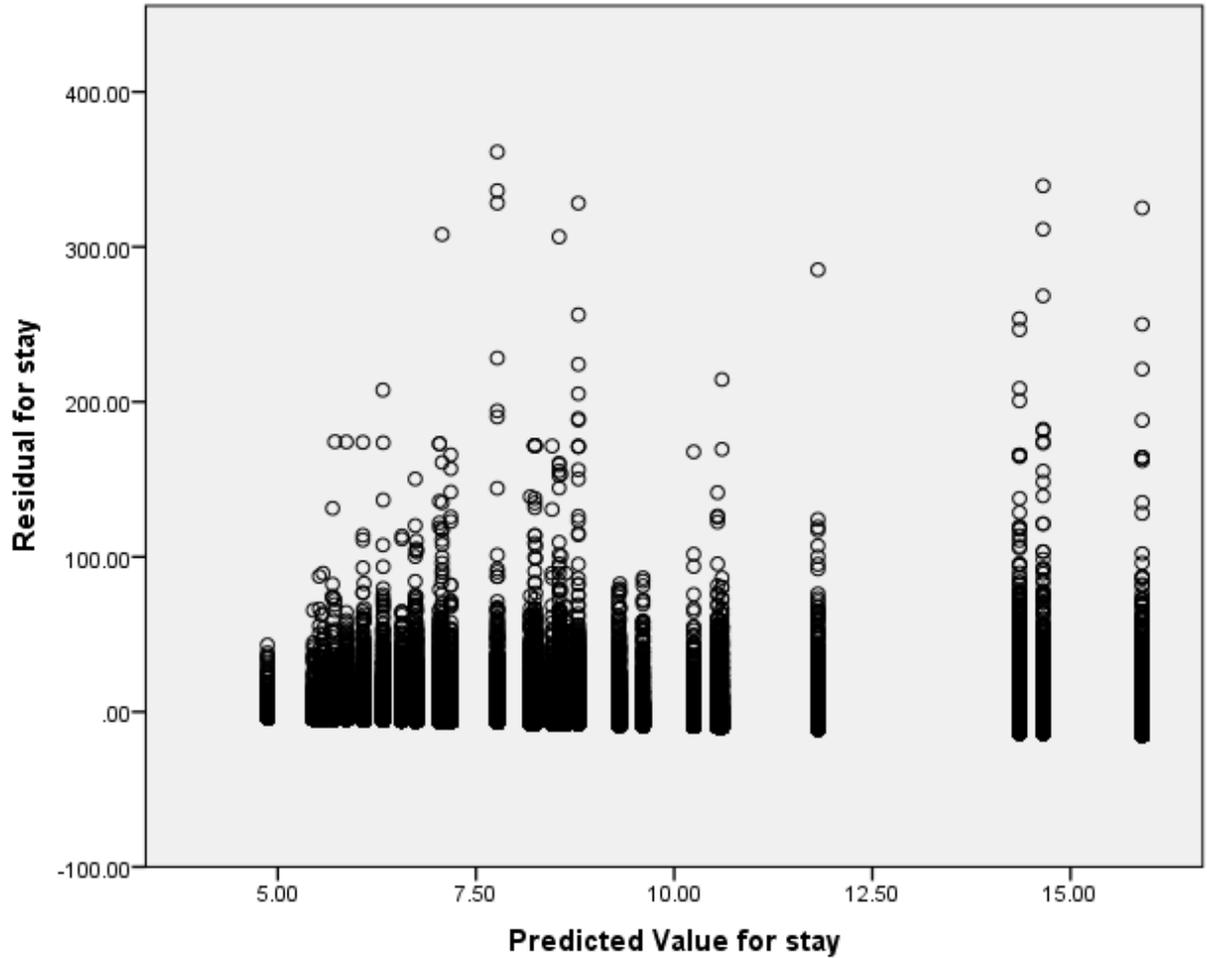

Figure 3. Plotting residual against predicted response variable

## 4.3. Adjusting the data

In skewed distributions, variance tends to be a function of the mean. If we can detect the relationship between the variance and mean, it will be possible to adjust the model, using variance stabilizing transformation. We have estimated the mean and standard deviation of the response variable for totally $ags = 5 * 2 * 4 = 40$ current cells, as $\hat{\sigma} = s_{ijk.}$ and $\hat{\mu} = \bar{y}_{ijk.}$. Fig. 4 plots $\log_{10} s_{ijk.}$ against $\log_{10} \bar{y}_{ijk.}$.

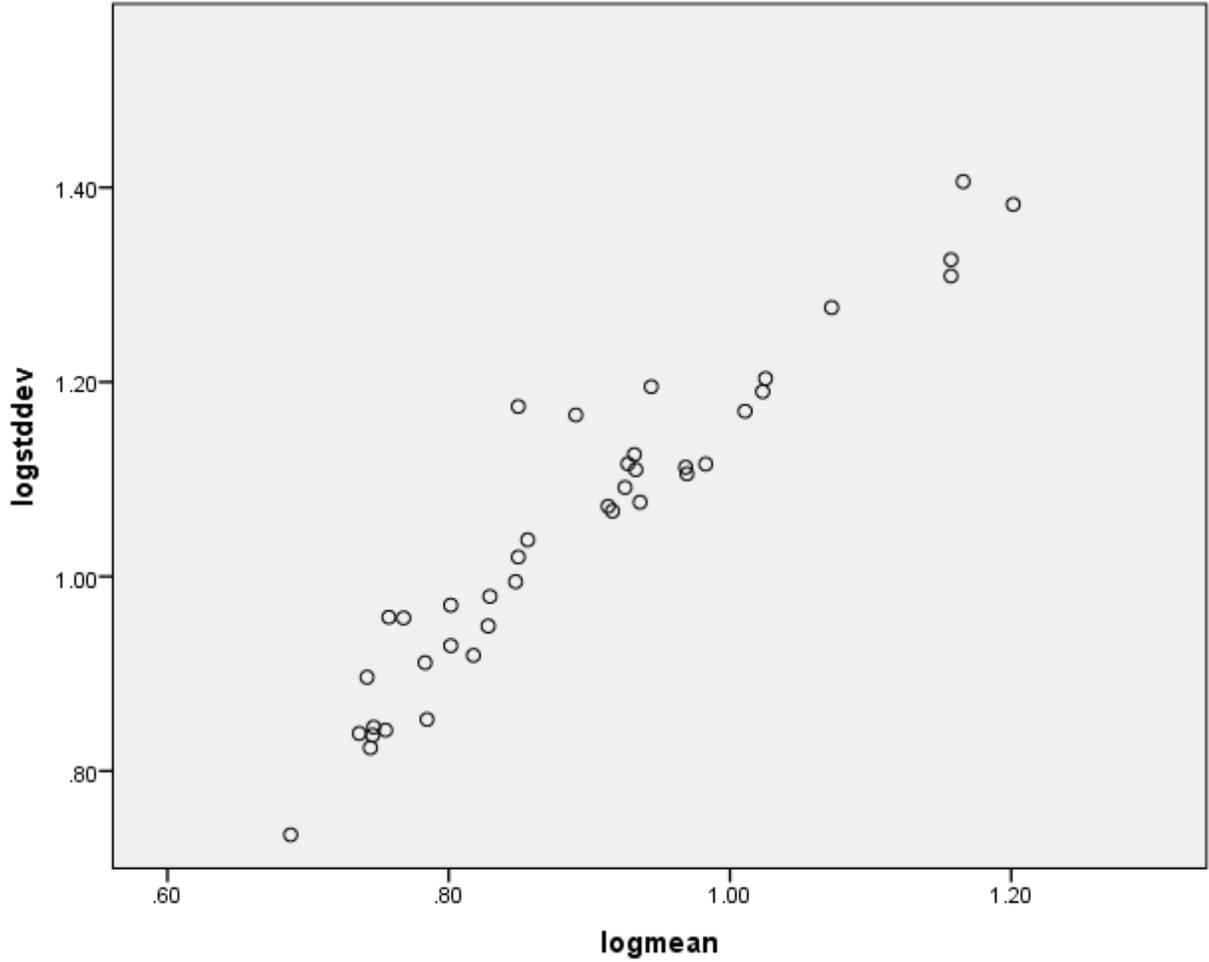

Figure 4. Plotting $\log_{10} s_{ijk.}$ against $\log_{10} \bar{y}_{ijk.}$

Fig. 4 signifies a linear relationship between two variables. Performing a regression analysis, we have a significant regression as Eq. 2 with $R^2 = 0.908$.

$$\log_{10} s_{ijk.} = 1.176 * \log_{10} \bar{y}_{ijk.} \tag{2}$$

Thus, the relationship between the mean and standard deviation of the response variable is estimated as Eq. 3.

$$\sigma = \mu^{1.176} \tag{3}$$

Montgomery [17] has introduced some variance stabilizing transformations for different exponents of µ. When the exponent nearly equals one, the proper transformation is logarithmic given as follows:

$$logstay = \log_{10}(LOS) \tag{4}$$

Having performed the variance analysis for the transformed data, we test the adjusted model's sufficiency, using the new residuals. Fig. 5 depicts the residuals' histogram for the adjusted model. In comparison to Fig. 2, the residuals' distribution is closer to normal and the results derived from this model are more reasonable.

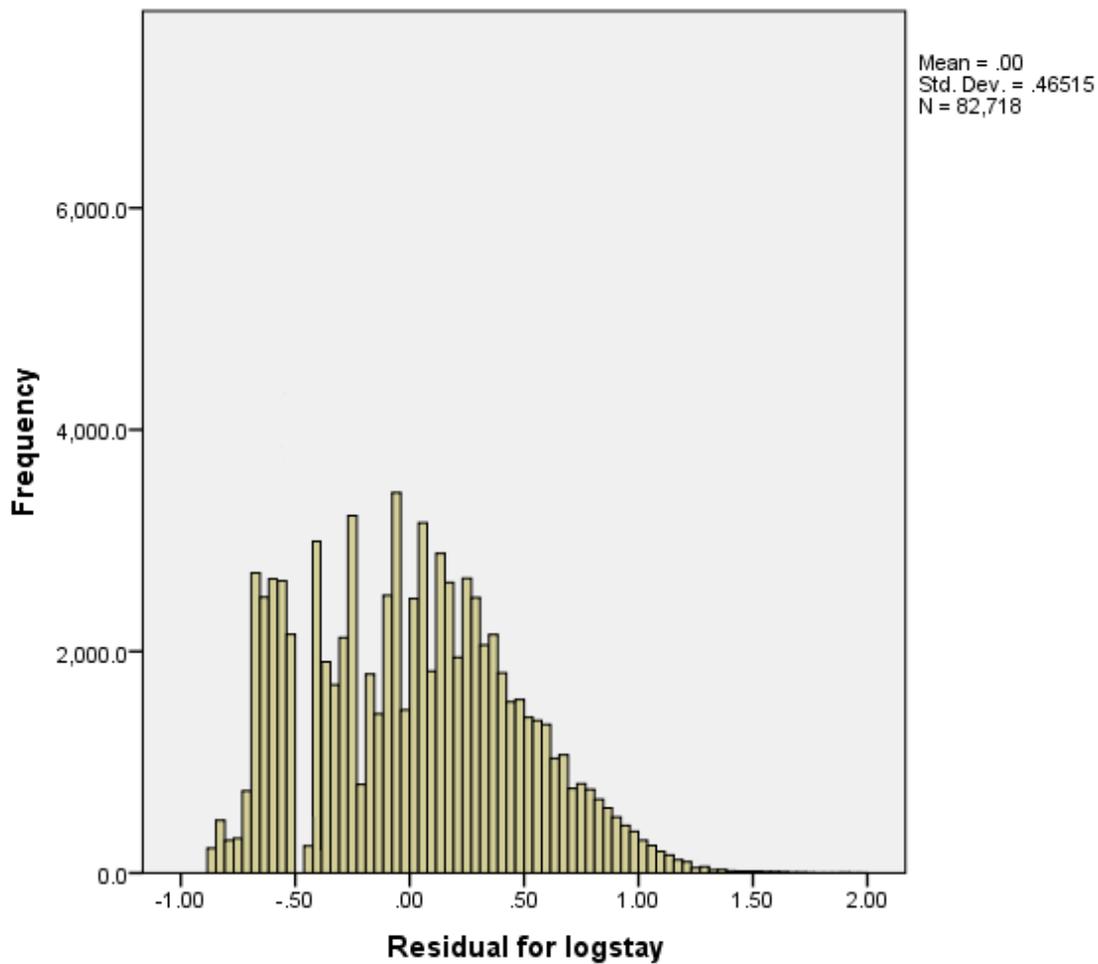

Figure 5. The residuals' histogram for the corrected model

Fig. 6 is normal P-P plot pertaining to new residuals. As the points are so close to the line, the normality assumption is not violated, however the nominal deviations can be observed.

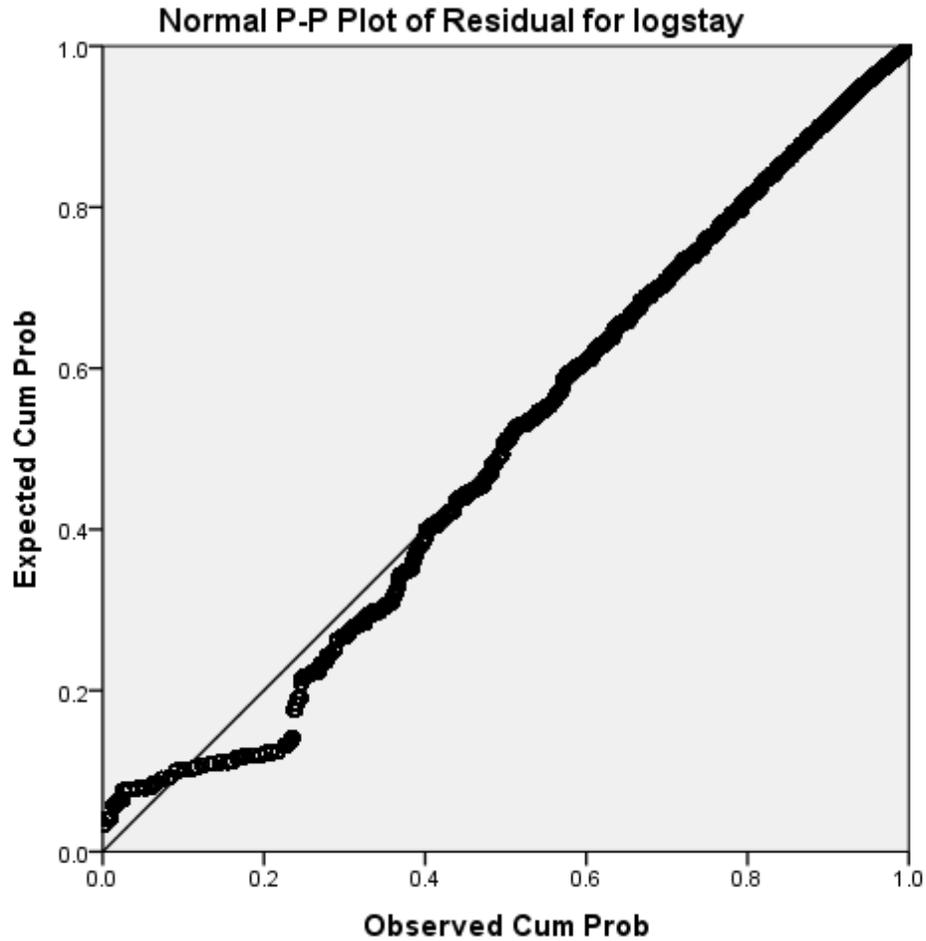

Figure 6. Normal P-P plot for residuals of the adjusted model

Finally, Fig. 7 represents the residuals against predicted values of the new response variable "logstay". In contrast to Fig. 3, this figure proves that the residuals have an approximately constant variance.

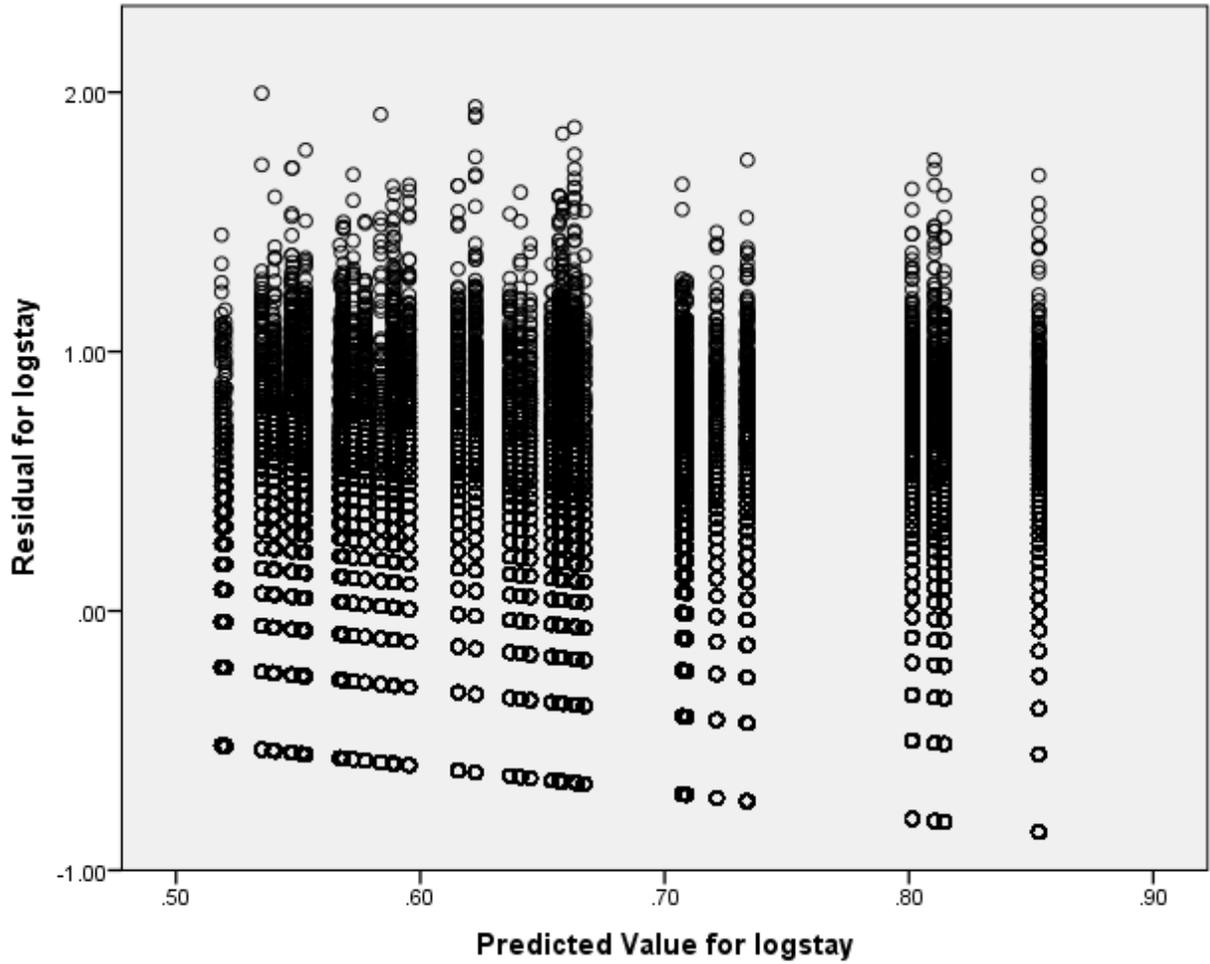

Figure 7. Plotting residuals against predicted values of the new response variable

## 5. Results

As the mentioned transformation has proved to be effective, we perform variance analysis for the transformed data. To observe the real values of the response variable, it's enough to calculate $LOS = 10^{\log stay}$.

### 5.1. Tests of between-subjects effects

The tests relating to significance of the whole model, the separate effects and interactions have been represented in Table 3.

Table 3. Tests of between subjects effects

| Source | Type III Sum of Squares | df | Mean Square | F | Sig. |
|---|---|---|---|---|---|
| Corrected Model | 513.847 | 39 | 13.176 | 60.866 | .000 |
| Intercept | 21676.552 | 1 | 21676.552 | 100137.437 | .000 |
| age group | 300.734 | 4 | 75.183 | 347.319 | .000 |
| season | 12.165 | 3 | 4.055 | 18.733 | .000 |
| gender | 46.741 | 1 | 46.741 | 215.926 | .000 |
| age group * season | 4.456 | 12 | .371 | 1.715 | .057 |
| age group * gender | 50.402 | 4 | 12.600 | 58.209 | .000 |
| season * gender | 1.783 | 3 | .594 | 2.745 | .041 |
| age group * season * gender | 2.354 | 12 | .196 | .906 | .540 |
| Error | 17897.142 | 82678 | .216 | | |
| Total | 50201.596 | 82718 | | | |
| Corrected Total | 18410.989 | 82717 | | | |

The results derived from this table are as follows:

- The whole model is significant ($P - value = 0.000 < 0.01$). It means at least one of the effects (separate effect or interaction) is significant.

- The separate effects of all three factors on the response variable are significant ($P-value = 0.000 < 0.01$). In other words, for each factor, the mean of logstay for at least two levels of the factor is not the same.
- Among interactions, only the one between two factors, say, "gender" and "age group" is significant ($P-value = 0.000 < 0.01$). It means that the value of the response variable at a level of each factor hinges on the level of the latter. The rest of interactions aren't significant at $\alpha = 0.01$, however the interaction of two factors "gender" and "season" is significant at $\alpha = 0.05$.

## 5.2. Estimating the model's parameters

The association between the factors and the response variable can be stated as a linear regression model. To do so, it is first needed to convert the factors into dummy variables. As an example, factor "season" with four levels changes to three dummy variables as follows:

$$season(1) = \begin{cases} 1 & spring \\ 0 & others \end{cases}$$

$$season(2) = \begin{cases} 1 & summer \\ 0 & others \end{cases}$$

$$season(3) = \begin{cases} 1 & autumn \\ 0 & others \end{cases}$$

Although this factor has four levels, to avoid collinearity, it is replaced with three dummy variables. Collinearity occurs when an independent variable can be written as a linear function of some other variables. The rest of dummy variables relating to other factors can be written similarly. Table 4 indicates the variables having significant effects on the response variable.

Table 4. Estimation of the regression model's parameters

| Parameter | B | Std. Error | t | Sig. | 95% Confidence Interval | |
|---|---|---|---|---|---|---|
| | | | | | Lower Bound | Upper Bound |
| Intercept | .573 | .008 | 69.655 | .000 | .556 | .589 |
| age group (2) | .161 | .019 | 8.453 | .000 | .124 | .198 |
| age group (3) | .091 | .016 | 5.754 | .000 | .060 | .123 |
| season (1) | -.037 | .011 | -3.407 | .001 | -.059 | -.016 |
| season (2) | -.032 | .011 | -2.805 | .005 | -.055 | -.010 |
| age group (2) * season (2) | -.056 | .026 | -2.172 | .030 | -.107 | -.006 |
| age group (3) * gender (1) | .133 | .021 | 6.333 | .000 | .092 | .175 |
| age group (4) * gender (1) | .053 | .016 | 3.289 | .001 | .021 | .084 |
| season (3) * gender (1) | .031 | .016 | 1.911 | .056 | -.001 | .063 |

Thus, considering $\alpha = 0.05$ the regression model is given as follows:

$$\log stay = 0.573 + 0.161[agegroup(2)] + 0.091[agegroup(3)] - 0.037[season(1)] - 0.032[season(2)]$$
$$-0.056[agegroup(2)*season(2)] + 0.133[agegroup(3)*gender(1)] + 0.053[agrgroup(4)*(gender(1)] \quad (5)$$

### 5.3. Post hoc tests

As the factors have proved to have significant separate effects, we are interested to know how the mean of the response variable changes over different levels of each factor.

### 5.3.1. comparing "logstay" means in different age groups

Different methods have been proposed to do post hoc tests. Among all these methods, Scheffe method is known as the most effective one to compare the means of groups with different sizes. This method is not sensitive to violation of basic assumptions, namely normality of residuals and variance homogeneity. However, in comparison to other methods, it seldom rejects the null hypothesis. Table 5 represents multiple comparisons of the response variable's means in different age groups. Only the first and fifth groups have close means (the means equality is rejected at $\alpha = 0.05$ but not rejected at $\alpha = 0.01$).

Table 5. Comparing "logstay" means in different age groups

| (I) age group | (J) age group | Mean Difference (I-J) | Std. Error | Sig. | 95% Confidence Interval | |
|---|---|---|---|---|---|---|
| | | | | | Lower Bound | Upper Bound |
| 1 | 2 | -.1594 | .00782 | .000 | -.1834 | -.1353 |
| | 3 | -.2020 | .00729 | .000 | -.2244 | -.1795 |
| | 4 | -.0708 | .00644 | .000 | -.0906 | -.0510 |
| | 5 | -.0199 | .00640 | .047 | -.0396 | -.0002 |
| 2 | 1 | .1594 | .00782 | .000 | .1353 | .1834 |
| | 3 | -.0426 | .00686 | .000 | -.0637 | -.0215 |
| | 4 | .0885 | .00594 | .000 | .0702 | .1068 |
| | 5 | .1395 | .00590 | .000 | .1213 | .1576 |
| 3 | 1 | .2020 | .00729 | .000 | .1795 | .2244 |
| | 2 | .0426 | .00686 | .000 | .0215 | .0637 |
| | 4 | .1311 | .00523 | .000 | .1150 | .1472 |
| | 5 | .1821 | .00519 | .000 | .1661 | .1981 |
| 4 | 1 | .0708 | .00644 | .000 | .0510 | .0906 |
| | 2 | -.0885 | .00594 | .000 | -.1068 | -.0702 |
| | 3 | -.1311 | .00523 | .000 | -.1472 | -.1150 |
| | 5 | .0509 | .00389 | .000 | .0390 | .0629 |
| 5 | 1 | .0199 | .00640 | .047 | .0002 | .0396 |
| | 2 | -.1395 | .00590 | .000 | -.1576 | -.1213 |
| | 3 | -.1821 | .00519 | .000 | -.1981 | -.1661 |
| | 4 | -.0509 | .00389 | .000 | -.0629 | -.0390 |

Table 6 indicates the subsets in which the response variable's means in different age groups do not significantly differ. As it is apparent no two age groups have the significant equal means and age groups 1 and 3 possess the minimum and maximum means of lgstay, respectively. Fig. 8 provides a visual explanation of the means for different age groups.

Table 6. The homogenous subsets for age groups

| age group | N | subset 1 | subset 2 | subset 3 | subset 4 | subset 5 |
|---|---|---|---|---|---|---|
| 1 | 6433 | 0.547 | | | | |
| 5 | 29456 | | 0.567 | | | |
| 4 | 27890 | | | 0.618 | | |
| 2 | 7875 | | | | 0.706 | |
| 3 | 11064 | | | | | 0.749 |
| sig. | | 1.000 | 1.000 | 1.000 | 1.000 | 1.000 |

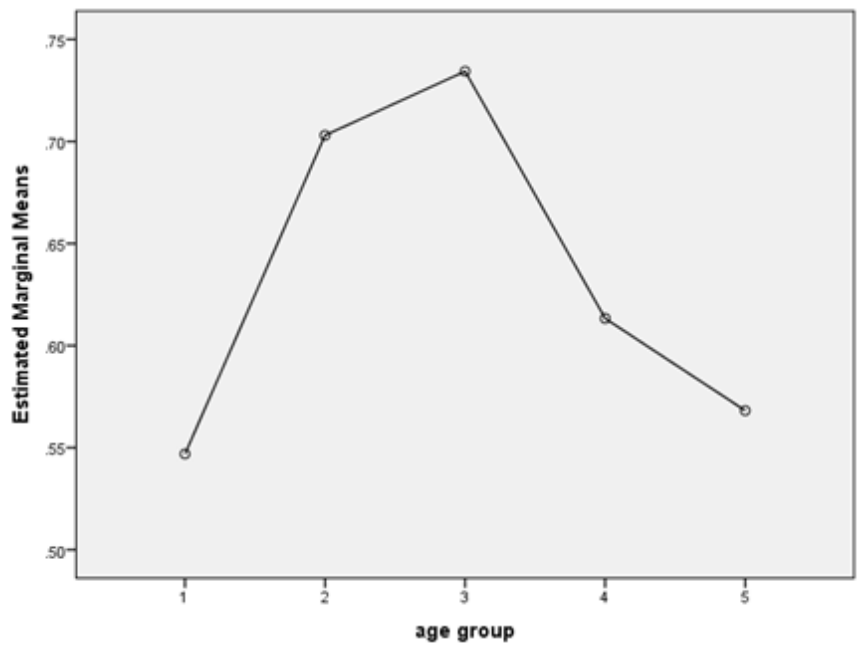

Figure 8. LOS variation over different age groups

## 5.3.2. Comparing "logstay" means in different seasons

As Table 7 explains, the only seasons in which the response variable's means do not significantly differ, are autumn and winter. The length of hospital stay in these two months are the highest, which might be due to the fact that cold weather enhances the prevalence of infectious maladies. Table 8 and Fig. 9 provide more wisdom on how the response variable's mean varies in different seasons.

Table 7. Comparing "logstay" means in different seasons

| (I) season | (J) season | Mean Difference (I-J) | Std. Error | Sig. | 95% Confidence Interval | |
|---|---|---|---|---|---|---|
| | | | | | Lower Bound | Upper Bound |
| spring | summer | -.0231 | .00446 | .000 | -.0356 | -.0106 |
| | autumn | -.0496 | .00459 | .000 | -.0625 | -.0368 |
| | winter | -.0407 | .00456 | .000 | -.0534 | -.0280 |
| summer | spring | .0231 | .00446 | .000 | .0106 | .0356 |
| | autumn | -.0265 | .00461 | .000 | -.0394 | -.0137 |
| | winter | -.0176 | .00458 | .002 | -.0304 | -.0048 |
| autumn | spring | .0496 | .00459 | .000 | .0368 | .0625 |
| | summer | .0265 | .00461 | .000 | .0137 | .0394 |
| | winter | .0089 | .00470 | .306 | -.0042 | .0221 |
| winter | spring | .0407 | .00456 | .000 | .0280 | .0534 |
| | summer | .0176 | .00458 | .002 | .0048 | .0304 |
| | autumn | -.0089 | .00470 | .306 | -.0221 | .0042 |

Table 8. The homogenous subsets for different seasons

|  | subset | | | |
|---|---|---|---|---|
| season | N | 1 | 2 | 3 |
| *spring* | 21963 | 0.593 | | |
| *summer* | 21546 | | 0.616 | |
| *winter* | 19817 | | | 0.633 |
| *autumn* | 19374 | | | 0.642 |
| *sig.* | | 1.000 | 1.000 | 1.000 |

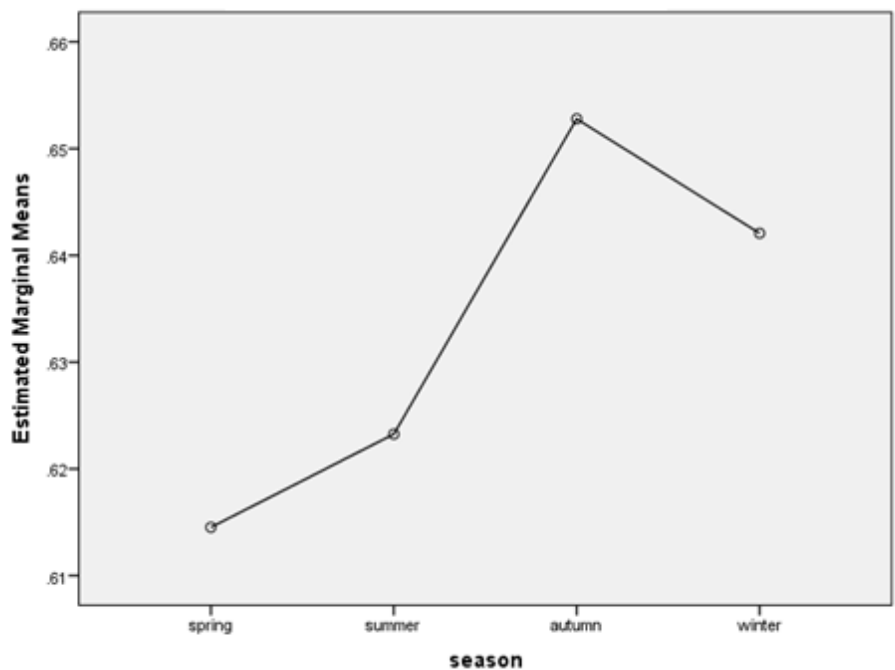

Figure 9. The homogenous subsets for different seasons

## 6. conclusion

In this research, we studied the effects of three factors, gender, season and age group on index LOS representing how long each patient stays in hospital, and proved the significance of these effects. Regarding the factor gender, men are hospitalized for longer periods, and length of stay enhances from spring to autumn and then descends in winter. However, autumn and winter approximately result in the equal length of stay. Among the different age groups, the patients aged 1 to 10 have the shortest stay and those at the age of 26 to 40 stay longer than the rest. Only two factors, gender and age group are proved to have significant interaction, meaning that the response variable's mean at a level of one factor hinges on the level of the latter. Considering these factors will help hospitals and healthcare centers to enhance healthcare service efficiency, allocate resources proportional to necessities, and remain within their budget. An advantage of the proposed model is the ease of determining associated factors' levels at a low cost. However, there might be some other factors directly or indirectly affecting length of the patients' stay, whose effects are yet to be well addressed.